\documentclass[a4paper]{article}

\usepackage{arxiv}
\usepackage{float}
\usepackage{graphics}
\usepackage{graphicx}
\usepackage{amssymb,amsmath,bm}
\usepackage{textcomp}
\usepackage{booktabs}
\usepackage[textfont=it,tableposition=top]{caption}
\usepackage{tikz}
\usepackage{pgfplots}
\usepackage[utf8]{inputenc} 
\usepackage[T1]{fontenc}    
\usepackage{hyperref}       
\usepackage{url}            
\usepackage{booktabs}       
\usepackage{amsfonts}       
\usepackage{nicefrac}       
\usepackage{microtype}      
\usepackage{cleveref}       
\usepackage{lipsum}         
\usepackage[square,sort,comma,numbers]{natbib}
\usepackage{doi}

\pgfplotsset{compat=1.17}

\title{R-MelNet: Reduced Mel-Spectral Modeling for Neural TTS}
\author{\hspace{1mm}Kyle Kastner \\
    Mila\\
    Universit\'e de Montr\'eal\\
	\And
	\hspace{1mm}Aaron Courville\thanks{$^\dagger$ CIFAR Fellow} \\
	Mila\\
	Universit\'e de Montr\'eal\\
}

\begin{document}

\maketitle
\begin{abstract}
  This paper introduces R-MelNet, a two-part autoregressive architecture with a frontend based on the first tier of MelNet and a backend WaveRNN-style audio decoder for neural text-to-speech synthesis. Taking as input a mixed sequence of characters and phonemes, with an optional audio priming sequence, this model produces low-resolution mel-spectral features which are interpolated and used by a WaveRNN decoder to produce an audio waveform. Coupled with half precision training, R-MelNet uses under 11 gigabytes of GPU memory on a single commodity GPU (NVIDIA 2080Ti). We detail a number of critical implementation details for stable half precision training, including an approximate, numerically stable mixture of logistics attention. Using a stochastic, multi-sample per step inference scheme, the resulting model generates highly varied audio, while enabling text and audio based controls to modify output waveforms. Qualitative and quantitative evaluations of an R-MelNet system trained on a single speaker TTS dataset demonstrate the effectiveness of our approach.
\end{abstract}

\section{Introduction}
Recent progress in text-to-speech (TTS) synthesis has been largely driven by the integration of neural networks which parameterize various factorizations of problems and subproblems in speech modeling \cite{black2007statistical}. This adoption has been driven by improved tooling \cite{paszke2019pytorch} as well as rapid improvement in stability and optimization for neural network training, allowing the development of ever larger and more complex TTS systems. Particularly, many recent neural TTS approaches feature a two-stage process, with one stage learning features from text-like inputs often in conjunction with a learned alignment procedure using attention modeling, or other language specific duration estimation models \cite{sotelo2017char2wav, wang2017tacotron, ping2018deep, chen2021wavegrad2}. 

Following this first stage model (often denoted as the \emph{frontend}), a secondary \emph{backend} network which focuses on upsampling, refining, or otherwise estimating audio waveforms from structured, time-aligned information is used to predict the final necessary output \cite{ kalchbrenner2018efficient, chen2020wavegrad}. Input information is generally a text-based representation, such as characters or phonemes, although recent models have also used self-supervised representations which combine audio and text information. Input modalities can be mixed at a per-datapoint level \cite{kastner2019representation, ping2018deep, fong2020testing}, in order to provide data-based regularization during training as well as the ability to flexibly choose between character and phonetic representations during inference.

The intermediate audio representation used for these two-stage systems is often a well-studied time-frequency representation from digital signal and speech processing, such as vocoder features, linear spectrograms, or log-mel spectrograms. These features can be coupled with high level information about delivery style, speaker characteristics, pitch curves, or any number of other high level controls useful for generative speech tasks.

The model types used for these various factorizations of the TTS problem come in a variety of forms: recurrent networks, convolutional networks, transformer models, or neural models combined with alternative probabilistic methods such as HMMs \cite{mehta2021neural}. One common variety of recurrent architecture builds on an attention based system for handwriting generation and text-to-vocoder synthesis \cite{graves2013generating}, using attention based recurrent networks to learn an alignment for turning character based inputs into audio outputs \cite{sotelo2017char2wav, wang2017tacotron, shen2018natural}. The basis for our frontend model, MelNet \cite{vasquez2019melnet}, builds upon this foundation as well as prior work on multi-dimensional recurrent models \cite{graves2008offline, theis2015generative, kalchbrenner2015grid, visin2015renet, van2016pixel}, in order to model time-frequency spectral features by jointly recurring over time and frequency.



\section{System Design}
\begin{figure}[!h]
\vspace{-1.0cm}
\begin{tikzpicture}
\node [anchor=west, rotate=-270] (freq) at (1.6,.8) {Mel-Freq (F)};
\node [anchor=west] (time) at (1.8,.6) {Time (T)};
\begin{scope}[xshift=0cm]
    \node[anchor=south west,inner sep=0cm] (image) at (0,0) {\includegraphics[width=0.9\columnwidth, height =120pt]{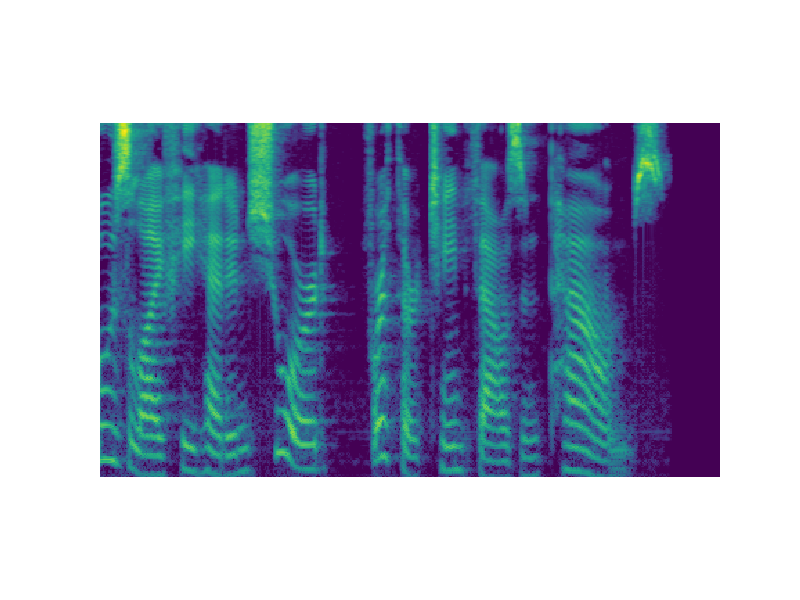}};
    \begin{scope}[x={(image.south east)},y={(image.north west)}]
        \draw [-latex, ultra thick, black] (freq) (0.105, 0.7) -- (0.105,0.8);
        \draw [-latex, ultra thick, black] (time) (0.225, 0.15) -- (0.9, 0.15);
    \end{scope}
\end{scope}
\end{tikzpicture}
\centering
\vspace{-.65cm}
\caption{Log-mel spectrogram (T=$448$, F=$256$).}
\label{fig:basemel}
\vspace{.2cm}
\hbox{\hspace{2.02cm}
\begin{tikzpicture}
\node [anchor=west, rotate=-270] (freq) at (-.2,-.11) {M-Fr};
\node [anchor=west] (time) at (0,-.24) {Time (T)};
\begin{scope}[xshift=0cm]
    \node[anchor=south west,inner sep=0cm] (image) at (0,0) {\includegraphics[width=0.7\columnwidth, height=30pt]{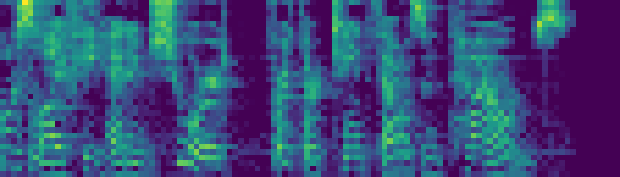}};
    \begin{scope}[x={(image.south east)},y={(image.north west)}]
        \draw [-latex, ultra thick, black] (freq) (-.017, 0.99) -- (-.017,1);
        \draw [-latex, ultra thick, black] (time) (0.13, -0.22) -- (1.0, -0.22);
    \end{scope}
\end{scope}
\end{tikzpicture}}
\vspace{-.2cm}
\caption{Downsampled log-mel spectrogram (T=$112$, F=$32$).}
\label{fig:downsampledmel}
\vspace{.25cm}
\end{figure}

The output target for the first tier of MelNet is a heavily downsampled log-mel spectrogram. In the case of a $5$ tier melnet using a downsampling factor of $2$ per tier (as used in R-MelNet), this results in data which is downsampled by $4$ in time, and $8$ in mel-frequency at the lowest level. As an example, we use a base log-mel spectrogram size (in time, frequency format) of ($448$, $256$), corresponding to approximately $4$ seconds of audio, as in Figure \ref{fig:basemel}. This leads to an overall log-mel spectrogram (in time, frequency format) of ($448$, $256$) reducing to a downsampled target of only ($112$, $32$) for the first tier loss, as in Figure \ref{fig:downsampledmel}. This extremely low-resolution data is predicted autoregressively (over both time and frequency), conditioned by attending over input text features. Time-frequency autoregression is handled by a structured arrangement of recurrent neural networks \cite{vasquez2019melnet}.

Subsequent tiers after the first in standard MelNet do not use an attention subnetwork, rather using the previously predicted, low-resolution log-mel spectrogram to condition the autoregressive generation (once again, over both time and frequency via recurrent networks) of a neighboring log-mel spectrogram - namely the log-mel spectrogram formed by downsampling in the same way but shifted by one column in time, or one row in frequency. Once predicted, the input conditioning and the predicted log-mel spectrogram are recombined by alternating either rows or columns into a map that is effectively twice as large in either time, or frequency. This means the overall tiers of the MelNet model a multi-scale sequence of log-mel spectrograms which gradually increase in resolution by $2$ in one-dimension at every tier after the first.

\subsection{A Dual-Purpose View of MelNet}
The iterative checkerboard of upsampling tiers in MelNet is reminiscent of several other image generative models \cite{dinh2016density, menick2018generating, saharia2021image}, and we can crudely break the overall MelNet architecture into two segments - the first tier, which predicts a low-resolution target conditioned only on text information using an attention subnetwork, and all subsequent tiers, which upsample the initial low-resolution predictions (and all intermediate predictions from previous tiers) into a high-quality, holistic log-mel spectrogram. Given this view of the overall MelNet architecture, and given the high computational cost of autoregressive sampling over larger and larger multi-dimensional maps, a natural question arises: \emph{Can we remove the upsampling tiers entirely, and opt instead to predict coarse, extremely low-resolution log-mel spectrograms which are paired with another inversion routine in order to obtain audio samples?}

Demonstrations of reduced tier samples from the original MelNet hint at the feasibility of this approach. In this work, we show that indeed, it is possible to remove all tiers except the first and instead use a WaveRNN to convert low-resolution first tier log-mel spectrograms into waveforms. Our model, denoted R-MelNet produces quality TTS with fewer hardware resources and parameters compared to the original implementation. 
\subsection{Architecture Details}
The R-MelNet frontend largely follows the recipe specified in MelNet, utilizing multi-dimensional recurrent neural networks, combined with an attention subnetwork over text features. Many of the details below are not specifically unique to R-MelNet, however we re-emphasize key aspects in order to clearly define our approach. 
The frontend network architecture uses $5$ combined layers, consisting of time, frequency, and centralized stacks \cite{vasquez2019melnet}. The attention network is inserted into the middle of the $5$ combined layers, conditioning all subsequent outputs, and all layers are connected with residual connections \cite{he2016deep, graves2013generating}.
We employ GRU units for all mel-spectrogram related recurrence, with hidden size for each GRU to $256$. For the attention subnetwork, we use a combination of embeddings to train with representation mixing \cite{kastner2019representation, ping2018deep, fong2020testing} at a $0.5$ per word swap probability between characters and phonemes, in order to enable either character-based or phoneme-based synthesis as well as providing data augmentation during training. The combined representation mixing embeddings are then fed directly into a bidirectional LSTM, forgoing the convolutional subnet which often links embeddings to the bidirectional LSTM in other work \cite{shen2018natural, kastner2019representation}. These resulting biLSTM hidden states form the memory which is attended over to condition the overall generation. All initial recurrent hidden states are fixed, starting from all zeros.
\subsection{Reduced Resource Requirements}
Neural networks are often limited by hardware resources, and GPU memory in particular is a frequent bottleneck for model training. We utilize several techniques to minimize memory requirements, keeping the overall memory necessary for training within $11$ gigabytes. All neural network code for these experiments was written using the PyTorch machine learning framework \cite{paszke2019pytorch}, and trained on a single 2080Ti NVIDIA GPU. One common way to reduce memory requirements is changing the floating point representation, from full precision ($32$ bit) to half precision ($16$ bit). This can bring a host of training complications due to numerical inaccuracy, but generally gives a substantial reduction in memory usage. We use half precision ($16$ bit) representations for all data and model parameters, as well as an Adam optimizer \cite{kingma2015adam} adapted for $16$ bit training \cite{brock2018large}. Despite these techniques, the memory required for moderately sized minibatches (on the order of $12$ to $16$) is still too large. One approach to further reduce memory use is to aggregate the gradient computation for a batch size $N$ over $K$ forward passes of sub-minibatches size $M$, where $M \times K = N$ \cite{hoffer2018augment}, thus allowing these \emph{virtual batches} to control memory usage independently of the effective batch size for optimization. We use a virtual batch size of $8$, with an effective batch size of $16$ in this work.
\subsection{Attention}
The attention subnetwork is a critical part of the overall architecture, as all text-based conditioning flows through this subnetwork, and the dynamics of the learned alignment define important aspects for TTS tasks. Attention calculations use many non-typical components and operations, and numerical precision is of critical importance. This is directly at odds with the half precision training strategy used in this work, and we made several modifications in order to ensure the stability of attention dynamics during training and sampling. 
R-MelNet uses a modified mixture of logistics attention \cite{vasquez2019melnet, battenberg2020location}, and we note that many of these changes may be necessary due to framework level implementation details between the PyTorch used here and the Tensorflow used in MelNet. We parameterize the mixture of logistics attention with $M$ mixture components
\begin{align}
    \{ \hat{\kappa}^{m}_i,\hat{\beta}^{m}_i,\hat{\alpha}^{m}_i \} &= W_g h_i \\
    \kappa^m_i &= \kappa^m_{i-1} + s_+(\hat{\kappa}^m_i) + \epsilon_{\kappa} \\
    \beta^m_i &= s_+(\hat{\beta}^m_i) \\
    \alpha^m_i &= \frac{\exp(\hat{\alpha}^m_i)}{\sum^M_{m=1}\exp(\hat{\alpha}^{m_i})}
\end{align}

Throughout the following equations, $\gamma_i$ is used as shorthand to represent the tuple of values ${\kappa_i,\beta_i,\alpha_i}$ parameterizing the location, scale, and weight for each mixture component. The term $\epsilon_{\kappa}$ is a small value to bound the minimum possible attention step, set to $.001$. Here $s_+$ represents a stable form of the softplus function 
\begin{equation}
    s_+(x) =\log(1 + \exp(-|x|)) + \max(0, x)
\end{equation}
Crucially, the $\log(1 + \exp(-|x|))$ function must be implemented in such a way that multiple numerical cases are handled in a branching fashion \cite{cook1log}, and that each branch of the computation which can result in not a number (NaN) values does not propagate this value outside the region of numerical stability for that branch. In PyTorch, this is accomplished by creating an output array, creating masks matching each branch condition, and filling subsets of the output array with the results of each branch, rather than directly performing a $where$ style conditional query.

This results in a final distribution function $F(u; \gamma_i)$ \cite{vasquez2019melnet}
\begin{align}
    F_i(u; \gamma_i) &= \sum^M_{m=1}\alpha^m_i(1 + \exp(\frac{\kappa^m_i - u}{\beta^m_i}))^{-1} \\
    \phi(u; \gamma_i) &= F_i(\mu + 0.5; \gamma_i) - F_i(u - 0.5; \gamma_i)
\end{align}

Here we note that $\phi(u; \gamma_i)$ can be parameterized using sigmoid calculations by expanding the difference and combining exponential terms. We use the equivalence between sigmoid and tanh to formulate a numerically similar alternative, using a cheap approximation to tanh and the equivalence of $\beta$ and $\frac{1}{\beta}$ when optimizing with the $s_+(x)$ function to avoid $\frac{1}{0}$ instability.

\begin{align}
    \tanh_{\sim} &= \frac{x}{1 + |x|} \\
    T(u; \gamma_i) &= \frac{1}{2} + \frac{1}{2}\tanh_{\sim}\left(\frac{(u - \kappa^m_i)(\beta^m_i + \epsilon_{\beta})}{2}\right) \\
    \phi(u; \gamma_i) &= \sum^M_{m=1} \alpha^m_i (T(u + 0.5; \gamma_i) - T(u - 0.5; \gamma_i))
\end{align}

The term $\epsilon_{\beta}$ is used to bound the minimum value, and is fixed to $.01$. Finally, we use a straight-through style gradient reweighting in order to use the direct value of $\phi(u; \gamma_i)$ in the forward pass, but reweight the gradient according to the number of attention components $\frac{1}{\sqrt{M}}$.
All intermediate attention values are calculated in full precision, and recast to half precision at the end of the attention weight calculation.

\subsection{The Importance of Inference Stochasticity}
The R-MelNet frontend uses mean squared error loss, as opposed to mixture density output and loss \cite{bishop1994mixture, graves2013generating, vasquez2019melnet}. Simple mean squared error, combined with numerical stability improvements, adaptive optimizers, and gradient clipping results in smooth, stable training. However when sampling, we note that the injection of stochasticity via noise is critical to achieve stable and high quality samples. 

We form an approximate distribution using $Q$ noise samples, by combining the average output mean predictions $\mu = \frac{1}{Q} \sum^Q_{q=1} \mu_q$ for a given model step with a particular standard deviation of noise, uniformly sampled over a predefined range $R$ to get $\sigma_r$ for this step. We next generate $Q$ samples from a truncated Gaussian distribution defined by this $(\mu, \sigma_r)$ pair, and these inputs feed the next model step. This gives the next set of mean predictions $\hat{\mu}^Q_{q=1}$, which are then averaged to form a new estimated mean $\hat{\mu}$, a new $\hat{\sigma_r}$ is drawn from range $R$, and so on. This procedure iterates for the total number of necessary sample steps to give a complete output. R-MelNet uses noise range $0 \leq R \leq 0.75$, and $1 \leq Q \leq 100$ noise samples, with the best settings at $.33$ and $100$ respectively. The result of sampling in this way can be seen in Figure \ref{fig:attention}.

This noise computation is easy to parallelize by permuting the differently noised inputs into the batch dimension, then averaging outputs along this dimension to estimate the average $\mu$. When utilizing this batch parallelism, the noisy sampling approach has relatively minor cost in terms of overall sampling speed.
The level of noise is critically important, as without noise or with few numbers of noise samples $Q$ the attention dynamics were highly unstable, and frequently led to issues with attention dynamics stalling as in Figure \ref{fig:attentionalt}. Higher noise range $R$ often resulted in more stable dynamics, but at the expense of final audio quality. Our results show that simple mean squared error losses can work when combined with noisy multi-sample inputs during sampling, though we recommend stochastic outputs (e.g. MDN) for future exploration.
Naive sampling schemes which recompute hidden states are extraordinarily costly in MelNet style multi-dimensional recurrent models. Sampling practically requires a fast caching/memoization scheme which caches all previous values for time delayed and centralized stacks, only recomputing when absolutely necessary because dependent variables change. This means that time delayed and centralized stacks only need to be recomputed when the frequency stack finishes an entire row of frequency predictions, while the frequency stack uses cached versions of the time and centralized stacks for intermediate computation.

\subsection{Framing the Context}
One advantage of MelNet style architectures is the ability to condition on past log-mel spectrogram information, in order to prime the style of the resulting audio. This gives the ability to generate varied outputs for a single text prompt by varying past mel-spectrum information and associated text, a critical aspect of stylistic and expressive TTS \cite{graves2013generating, chang2021style, stanton2021speaker}. This ability to prime the model with audio context, coupled with the ability to granularly control text information on a per-word basis using representation mixing, means it is possible to generate many possible candidate samples for a single desired phrase. Though this is advantageous for creating variable output, it also makes the selection of a single best model output from a collection of different priming audio and text sequences into a complex search problem. 

For fully automated TTS, we use a simple heuristic search algorithm to refine and rerank proposed model samples, based on the smoothness, continuity, and completion of the attention weights. See Figure \ref{fig:attentionalt} for an example. After throwing out attentions which failed to reach the end of the text, or broke apart (detected by contiguity of a mask based on the $95$th percentile value), our ranking function looks at the nearness to the median length (across all seeds which succeeded), gap between the max and median attention trace value, gap between the max and min attention trace value, and the reciprocal of the min attention trace value. These four values are multiplied together, and the results ranked from lowest to highest, choosing the result with the lowest value. This means that attentions with a high constant value on the attention trace, that aren't unusually short or long compared to other seeds should get a low score, whereas uncertain or highly variable attentions, or ones with an especially fast or slow rate will get a higher score. This selection method could integrate automatic metrics as a direction for future work \cite{huang2022voicemos}.

\subsection{Converting Low-Resolution Mel-Spectrograms into Audio}
\begin{figure}[!tbp]
\centering
\includegraphics[width=.8\columnwidth,height=60pt]{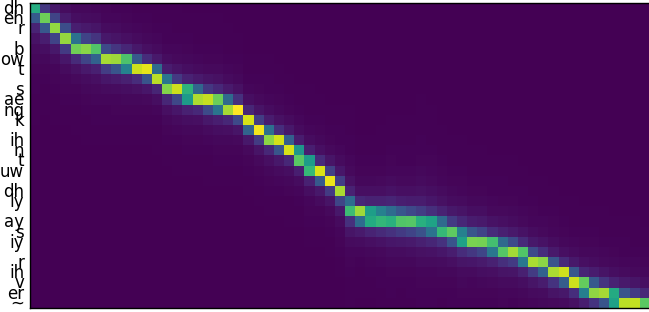}
\hbox{\hspace{.3cm}\includegraphics[width=.765\columnwidth,height=60pt]{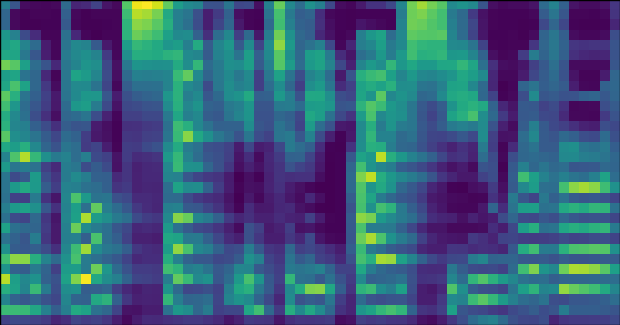}}
\hbox{\hspace{.3cm}\includegraphics[width=.765\columnwidth,height=60pt]{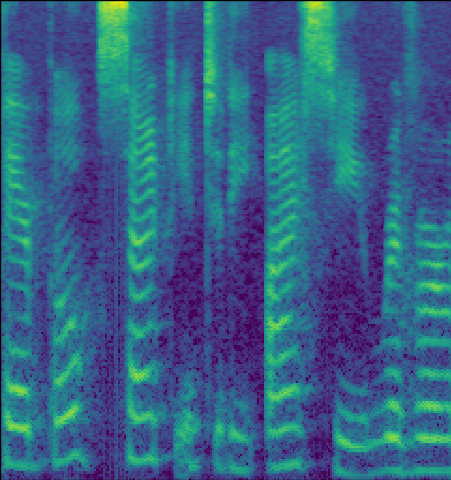}}
\caption{Attention alignment and sampled output from Mel-Net frontend along with log-mel spectrogram of final output waveform from WaveRNN backend, for example input 'their boat sank into the icy river', phonemized to 'dhehr bowt saengk ihntuw dhiy aysiy rihver'.}
\label{fig:attention}
\end{figure}

\begin{figure}[!tbp]
\centering
\includegraphics[width=.3\columnwidth, height=40pt]{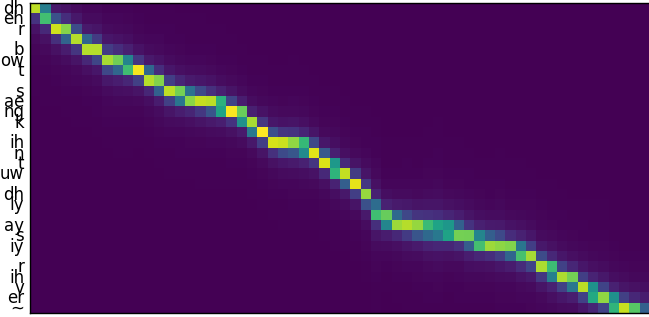}
\includegraphics[width=.3\columnwidth,height=40pt]{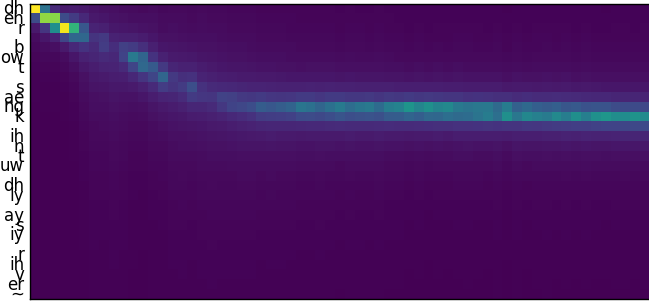}
\includegraphics[width=.3\columnwidth,height=40pt]{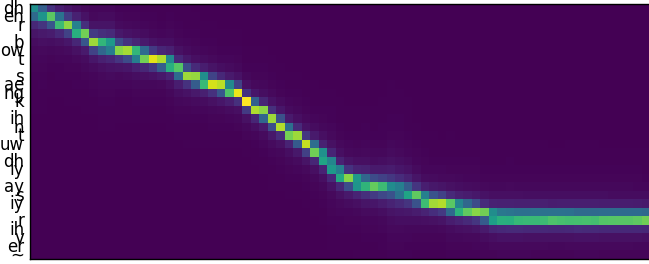}
\caption{Example of alternate successful attention alignment, and two failed alignments for the same sentence as Figure \ref{fig:attention}, starting with different audio and text priming.}
\label{fig:attentionalt}
\end{figure}

As described above, predicting low-resolution log-mel spectrograms from R-MelNet is the first stage of the overall audio generation. Once the low-resolution output is predicted, another stage is necessary to turn the predicted output (see Figure \ref{fig:attention}) into listenable audio. We use a WaveRNN \cite{kalchbrenner2018efficient} for this purpose, in order to preserve the autoregressive nature of audio decoding. Our setup follows the normal procedure for mel to audio WaveRNN training \cite{mccarthy1wavernn}, with one crucial difference. We found that modifying the convolutional encoder of WaveRNN to upsample the time dimension of the $(112, 32)$ input by $4$, back to the native time resolution performed poorly. Instead, it was beneficial to linearly interpolate (per frequency bin) the low resolution input up to $(448, 32)$ before input to the WaveRNN pipeline. Many models \cite{chen2020wavegrad, lam2021bilateral, morrison2021chunked} can be used for this mel-to-audio stage. Exploring which audio models best predict high-quality audio data given low-resolution input is an important direction for future work, as well as analyzing sources of pitch variation and pitch error \cite{morrison2021chunked, turian2020m}.

\section{Experiments}
\label{section:experiments}
Given the massive number of components to this system, we encourage readers to refer to the experimental code \footnote{\url{https://github.com/kastnerkyle/rmelnet}} for hyperparameters and other key implementation details. All training for R-MelNet, as well as comparison baselines Fastspeech 2 and Portaspeech, used the LJSpeech dataset \cite{ljspeech17}. We note that most audio preprocessing settings for R-MelNet follow precedent work \cite{yamamoto1deepvoice}, with base audio at $22.050$ kHz sample rate, using STFT size $6 \times 200$, STFT step $200$, and base mels $256$ before downsampling. We also found global mean and std normalization per frequency bin for the R-MelNet frontend was crucial for stable training.

\begin{table}[ht]
\centering
\begin{tabular}{|c||c|c|}
\hline
Model & MOS-P (95\% CI) & Parameters \\
\hline
\hline
R-MelNet & $3.686 \pm 0.144$ & 43M \\
\hline
Fastspeech 2 \cite{wang2020fairseq} & $3.6736 \pm 0.151$ & 27M \\
\hline
Portaspeech \cite{ren2021portaspeech} & $3.278 \pm 0.183$ & 21.8M \\
\hline
\end{tabular}
\vspace{.2cm}
\caption{MOS-P listening study results, comparing Fastspeech 2, Portaspeech, and R-MelNet.}
\label{table:mosp}
\vspace{-.2cm}
\end{table}

R-MelNet performs slightly better on average in a prosody-focused mean opinion score (MOS-P) test compared to Fastspeech 2 (as implemented by fairseq $S^2$, with HiFi-GAN vocoder \cite{kong2020hifi}) and Portaspeech (also with HiFi-GAN vocoder), though confidence intervals for R-MelNet and Fastspeech 2 are highly overlapped. Test participants were asked to ignore small variations in audio quality, so although the resulting R-MelNet audio is slightly noisier than the equivalent outputs from Fastspeech 2 and Portaspeech, it did not overly affect ratings. In a quality based MOS test (MOS-Q) we would expect to see lower performance than these baselines, but pure audio quality was not the focus of our current work. We use $14$ sentences from the Harvard Sentences \cite{ieeetra69}, along with $11$ custom sentences for a total of $25$ sentences, which were synthesized using all $3$ models, and presented in random order to each user in the test. Given the ratings for this test ($261$ for each model, across $34$ separate users), and overlap in confidence intervals, it is difficult to discern an absolute ranking of performance. We do not take these scores as a critique of the comparison methods, only as a validation that R-MelNet is of comparable quality to both Fastspeech 2 and Portaspeech, proving the validity of our approach for neural TTS.
The spread of options for sampling and high variability in R-MelNet are potentially suited to interactive use, however in the current implementation sampling speed ($\sim4$x slower than real-time) and frame latency ($\sim50$ ms per mel frame) are not fast enough for human-in-the-loop usage. An ideal, pipelined version of R-MelNet could meet a $30$ ms latency, sub real-time output threshold which would enable interactive use, but requires significant engineering effort. As such we leave this exploration to future work.



\section{Conclusions}
We describe R-MelNet, a two-stage architecture for neural text-to-speech synthesis. The R-MelNet frontend uses a one tier version of MelNet, which takes in a mixed sequence of character and phoneme features, along with an optional audio priming sequence, and outputs low-resolution mel-spectral features. These features are in turn upsampled, and input to a WaveRNN which generates an audio waveform.
The audio from this generative pipeline is varied and expressive, giving many possible outputs for a single desired prompt. By incorporating practical methods for reducing computational complexity and memory usage such as half precision training with numerically stable implementations of core operations, our method enables training on a single commodity gpu.
This resulting model compares favorably in terms of parameter count, 
and quality with recent non-autoregressive TTS methods when trained on an openly available single speaker dataset.

\section{Acknowledgements}
The authors would like to thank Tim Cooijmans, Cheng-Zhi Anna Huang, Yusong Wu, Laurent Dinh, and Johanna Hansen for helpful feedback on versions of this work. The authors would also like to thank all participants in our listening tests.
\newpage
\bibliographystyle{IEEEtran}

\bibliography{mybib}


\end{document}